\documentclass[aps,preprint,superscriptaddress,nofootinbib]{revtex4-1}   
\pdfoutput=1
\usepackage[utf8]{inputenc}
\usepackage{amssymb}
\usepackage{amsmath}
\usepackage{amsfonts}
\usepackage{graphicx}
\usepackage{color}
\usepackage{xspace}
\usepackage{comment}
\usepackage{hyperref}
\usepackage[normalem]{ulem}

\usepackage[section]{placeins}
\usepackage{afterpage}

\usepackage{float}
\usepackage{slashed}
\usepackage{ulem}
\usepackage{appendix}

\usepackage{cancel}

\usepackage{multirow,rotating}
\usepackage[dvipsnames]{xcolor}
\usepackage{orcidlink}   

\newcommand{\bea} {\begin{eqnarray}}
\newcommand{\eea} {\end{eqnarray}}

\newcommand{\beq} {\begin{equation}}
\newcommand{\eeq} {\end{equation}}
\newcommand{\order}{\mathcal{O}}

\begin{document}
	
      \title{Probing pair production of long-lived scalars via an off-shell Standard-Model-like Higgs boson at the LHC}

      \author{Lei Wang\,\orcidlink{0000-0001-8639-4917}}
      \affiliation{Department of Physics, Yantai University, Yantai
264005, China}

      \author{Xianbo Yu}
      \affiliation{Department of Physics, Yantai University, Yantai
264005, China}

	   \author{Zeren Simon Wang\,\orcidlink{0000-0002-1483-6314}}
	   \email{wzs@hfut.edu.cn}
      \affiliation{School of Physics, Hefei University of Technology, Hefei 230601, People’s Republic of China}

\begin{abstract}
We study the collider phenomenology of a long-lived scalar particle $S$ that arises from Higgs mixing in a broad class of Standard-Model (SM) extensions. When the mixing angle is sufficiently small, $S$ becomes long-lived, while its pair production via the Higgs portal can remain sizable. We focus on the production channel $gg \to h^* \to SS$ at the LHC, mediated by an \textit{off-shell} SM-like Higgs boson. This mechanism provides a complementary probe of $S$ in the mass region above the kinematic threshold of the conventional on-shell decay $h \to SS$, thereby extending the accessible parameter space to heavier scalars. The long-lived $S$ particles can decay inside the inner detector, leading to displaced vertices (DVs) accompanied by jets. We perform a detailed Monte Carlo simulation and reinterpret an existing recast of an ATLAS search for DV-plus-jets signatures in this scenario. We also consider a modified analysis strategy based on the same search to assess potential improvements in sensitivity. We find that the current ATLAS search already excludes a significant region of the parameter space, reaching scalar masses up to $m_S \sim 230$~GeV for a benchmark $hSS$ coupling $\lambda v$ of $246$~GeV. The modified analysis and projections to the high-luminosity LHC further extend the sensitivity to wider regions of the mass--lifetime parameter space.
\end{abstract}



	\maketitle
    \noindent

\section{introduction}\label{sec:intro}

Since the discovery of the Higgs boson, increasingly precise measurements at the LHC have shown that its properties are largely consistent with the predictions of the Standard Model (SM)~\cite{ATLAS:2012yve,CMS:2012qbp}.
Despite this success, the possibility of an extended scalar sector remains well motivated, and searching for additional scalar states continues to be an important goal in collider physics.

A particularly intriguing possibility is that new scalar particles are long-lived.
Such states can naturally arise in a variety of SM extensions, such as the real-scalar extension of SM~\cite{Liu:2022nvk}, specific realizations of two-Higgs-doublet models (2HDMs)~\cite{Wang:2022yhm,Kling:2022uzy,Haisch:2023rqs,Liu:2024azc,Shan:2024pcc,Kim:2025tuz,Qi:2025qsj,Wang:2025zhd}, Georgi-Machacek model \cite{Lu:2024ade,Fang:2025iui}, neutral-naturalness models~\cite{Burdman:2006tz,Chacko:2005pe,Craig:2015pha,Cai:2008au,Cohen:2018mgv,Cheng:2018gvu}, and non-minimal supersymmetric frameworks~\cite{Ellwanger:2009dp,Cao:2013gba,Adhikary:2022pni}.
While ATLAS and CMS have performed extensive searches for new scalar states, most analyses have focused on prompt signatures~\cite{ATLAS:2020ahi,ATLAS:2024vxm,CMS:2024zfv,CMS:2023boe}, despite an increasing number of reported searches for non-prompt signatures~\cite{ATLAS:2019qrr,ATLAS:2025pak,CMS:2024xzb,CMS:2024bvl}.
Indeed, in recent years, more and more attention has been devoted to the scalar as a long-lived particle (LLP).\footnote{See, e.g., Refs.~\cite{Alimena:2019zri,Lee:2018pag,Curtin:2018mvb,Beacham:2019nyx,Jeanty:2025wai,Alimena:2025kjv} for some recent reviews on LLPs.}

From both theoretical and experimental perspectives, a widely studied production mechanism for long-lived scalars is through the decay of an on-shell SM-like Higgs boson ($h$), 
$h\to SS$, where $S$ denotes a long-lived scalar particle~ (see, e.g., Refs.~\cite{Alipour-Fard:2018lsf,Filimonova:2019tuy,Cheung:2019qdr,Liu:2022nvk,Wang:2024ieo,Cepeda:2021rql}). However, this production mode is kinematically limited to scalar masses below $m_h/2$,  thereby restricting its sensitivity to relatively light long-lived states. Although above the kinematic threshold, the on-shell SM-like Higgs boson can in principle decay into one on-shell $S$ and one off-shell $S$, the corresponding branching ratio is extremely suppressed by the tiny interactions of the long-lived scalar $S$, and is therefore not suitable for probing $S$.

In this work, we explore an alternative production mechanism in which the intermediate SM-like Higgs boson is off shell, namely through the process $gg\to h^* \to SS$.
This mechanism provides a complementary probe of a long-lived scalar in the mass region above $m_h/2$, significantly extending the accessible parameter space.
We focus on a class of new-physics (NP) scenarios in which the scalar $S$ acquires its couplings to the SM fermions and gauge bosons through mixing with $h$. As a result, when the mixing angle is sufficiently small, $S$ can become long-lived.
However, such small mixing angle can not affect the $hSS$ coupling, allowing for sizable production rates via the off-shell $h$ mediator.
Depending its mass, the scalar $S$ predominantly decays into $b\bar{b}$ or $WW$.
If the scalar $S$ travels a macroscopic distance before decaying inside the inner detector, the resulting signal is characterized by displaced vertices (DVs) plus jets at the ATLAS or CMS detectors.

In this work, considering this signature, we apply an existing recast~\cite{Cheung:2024qve} of an ATLAS search for DVs and multiple jets~\cite{ATLAS:2023oti} and reinterpret its exclusion limits in terms of the scenario of long-lived scalars from off-shell Higgs decays.
We further consider the search prospects of an analysis modified from this search and those at the high-luminosity LHC (HL-LHC).

The paper is organized as follows. In Sec.~\ref{sec:model}, we introduce the theoretical framework in which a long-lived scalar $S$ mixes with the SM-like Higgs boson, including two representative NP models. 
In Sec.~\ref{sec:experiment}, we describe the experimental search at ATLAS, its existing recast, as well as a modified analysis.
Then in Sec.~\ref{sec:results} we present and discuss the corresponding numerical results.
Finally, we summarize our conclusions in Sec.~\ref{sec:summary}.

\section{A long-lived scalar mixed with the SM-like Higgs boson}\label{sec:model}
\label{sec:model}
We consider a class of NP scenarios that share the following phenomenological form of the interactions of the scalar $S$,
\begin{align}
\mathcal{L}_{S}  \supset \frac{m_f}{v}\sin\theta \bar{f}fS-2\frac{m^2_W}{v}\sin\theta W_{\mu}^+W^{-\mu} S -\frac{m^2_Z}{v}\sin\theta Z_{\mu}Z^{\mu} S - \frac{\lambda}{2} v hSS+\cdots,
\label{effcoup}
\end{align}
where $v=246$ GeV, $\theta$ is the mixing angle between $S$ and $h$, and $f$ denotes the charged leptons and quarks. 
In the following, we introduce two representative NP ultraviolet-complete scenarios.

\subsection{Real-singlet-scalar extension of the SM}
The scalar potential in the SM extended by a real singlet scalar field can be written as
\begin{eqnarray} \label{VSMS} &&\mathrm{V} =-\mu^2
(\Phi^{\dagger} \Phi) + \frac{\lambda_1}{2}  (\Phi^{\dagger} \Phi)^2 +
\frac{\mu_S^2}{2} \mathbb{S}^2 + \frac{\lambda_2}{8}  \mathbb{S}^4 +
\frac{\lambda_3}{3} \mathbb{S}^3 + \frac{\lambda_4}{2} \Phi^{\dagger} \Phi \mathbb{S} +  \frac{\lambda_5}{2} \Phi^{\dagger} \Phi \mathbb{S}^2,
\end{eqnarray}
where a tadpole term is removed by the shift invariance of the singlet potential.
$\Phi$ corresponds to the SM Higgs doublet field and $\mathbb{S}$ is the scalar singlet field, which have non-zero vacuum expectation values (VEVs),
\bea
&&\Phi=\left(\begin{array}{c} \phi^+ \\
\frac{(v+\phi^0+i\eta)}{\sqrt{2}}\,
\end{array}\right)\,, ~~~~~
\mathbb{S}=v_s+s.
\eea
The physical scalar states arise from the mixing of $\phi^0$ and $s$,
\begin{eqnarray}
\left(\begin{array}{c}h \\ S \end{array}\right) =  \left(\begin{array}{cc}\cos\theta & \sin\theta \\ -\sin\theta & \cos\theta \end{array}\right)  \left(\begin{array}{c} \phi^0 \\ s \end{array}\right),
\end{eqnarray}
and such mixing leads $S$ to have the couplings to fermions and gauge bosons as shown in Eq.~\eqref{effcoup}.
The cubic scalar interactions between $h$ and $S$ contains
\beq
\mathcal{L}_{\text{cubic}} = - \lambda_{hSS} hSS - \lambda_{Shh} Shh,
\eeq
with
\bea
\lambda_{hSS} &=& \frac{\lambda_5 v c_\theta^3  + (3\lambda_2 v_s -2 \lambda_5 v_s + 2\lambda_3 -\lambda_4)  c_\theta^2 s_\theta}{2}, \nonumber\\
\lambda_{Shh} &=& \frac{(\lambda_4 +2 \lambda_5 v_s) c_\theta^3  + (4 \lambda_5 - 6\lambda_1 ) v c_\theta^2 s_\theta}{4}.
\eea
Here, we retain terms only up to $\order(s_\theta)$. 

The scalar mixing angle is related to the parameters of the scalar potential through
\beq
\lambda_4=-2\lambda_5 v_s+\frac{2(m_h^2-m_S^2) s_\theta c_\theta}{v}.
\eeq
For very small values of $s_\theta$, the second term is highly suppressed.
Therefore, if $\lambda_5=\mathcal{O}(1)$ and $v_s$ is not small, achieving $s_\theta\sim10^{-8}$--$10^{-6}$ requires a significant cancellation between $\lambda_4$ and $2\lambda_5 v_s$.
Such a cancellation is not protected by an approximate symmetry and may be regarded as a fine-tuned realization of the model.
The small mixing angle is thus simply treated as a free parameter in the present phenomenological analysis.

It should be emphasized that arbitrary choices of $v_s$, $\lambda_5$, and tiny  $s_\theta$ do not necessarily correspond to a theoretically consistent scalar potential.
These parameters are subject to the minimization conditions of the scalar potential and the requirement that the electroweak vacuum be the desired global minimum.

\subsection{Next-to-two-Higgs-doublet model}
The next-to-two-Higgs-doublet model (N2HDM) is a well-motivated extension of the SM in which a real singlet scalar field is added to the conventional 2HDM~\cite{Chen:2013jvg,Muhlleitner:2016mzt}.
The Higgs potential is written as
\begin{equation}
  \begin{aligned}
  V &= m_{11}^2 |\Phi_1|^2 + m_{22}^2 |\Phi_2|^2 - m_{12}^2 (\Phi_1^\dagger
  \Phi_2 + \text{h.c.}) + \frac{\lambda_1}{2} (\Phi_1^\dagger \Phi_1)^2 +
  \frac{\lambda_2}{2} (\Phi_2^\dagger \Phi_2)^2  \\
  &\hphantom{=} + \lambda_3
  (\Phi_1^\dagger \Phi_1) (\Phi_2^\dagger \Phi_2) + \lambda_4
  (\Phi_1^\dagger \Phi_2) (\Phi_2^\dagger \Phi_1) + \frac{\lambda_5}{2}
  [(\Phi_1^\dagger \Phi_2)^2 + \text{h.c.}]  \\
  &\hphantom{=} + \frac{m_S^2}{2}  \Phi_S^2 + \frac{\lambda_6}{8} \Phi_S^4 +
  \frac{\lambda_7}{2} (\Phi_1^\dagger \Phi_1) \Phi_S^2 +
  \frac{\lambda_8}{2} (\Phi_2^\dagger \Phi_2) \Phi_S^2,
\end{aligned}
  \label{eq:n2hdmpot}
\end{equation}
with the VEVs given by
\bea
&&\Phi_1=\left(\begin{array}{c} \phi_1^+ \\
\frac{(v_1+\rho_1+i\eta_1)}{\sqrt{2}}\,
\end{array}\right)\,, 
\Phi_2=\left(\begin{array}{c} \phi_2^+ \\
\frac{(v_2+\rho_2+i\eta_2)}{\sqrt{2}}\,
\end{array}\right),
\Phi_S=\rho_s+v_s.
\eea
After spontaneous symmetry breaking, the scalar fields $\Phi_1, \Phi_2$, and $\Phi_S$ develop VEVs
 $v_1$, $v_2$, and $v_s$, respectively, with $v\equiv \sqrt{v_1^2+v_2^2}=246$ GeV. The ratio of the two-Higgs-doublet VEVs is conventionally expressed as 
 $\tan\beta \equiv v_2 /v_1$. 

The physical three CP-even scalar states are obtained from the original fields via a rotation matrix,
\begin{eqnarray}
\left(\begin{array}{c}S,h,H \end{array}\right) =   \left(\begin{array}{c} \rho_1,  \rho_2,  \rho_s\end{array} \right) R^T, 
\end{eqnarray}
with 
\beq
	R= \left( \begin{array}{*{20}{c}}
			c_{1}   c_{2} & s_{1}  c_{2}  & s_{2}\\
			  -  s_{1}  c_{3}-c_{1}   s_{2}   s_{3}  &  c_{1}  c_{3}-s_{1}  s_{2}  s_{3}   & c_{2}  s_{3}  \\
			 s_{ 1}  s_{3}-c_{1}  s_{2}  c_{3} &- s_{1}  s_{2}c_{3}  -c_{1}  s_{3}  & c_{2}  c_{3}  
	\end{array} \right).
\eeq
The shorthand notations are $s_{1,2,3}\equiv\sin\alpha_{1,2,3}$ and $c_{1,2,3}\equiv\cos\alpha_{1,2,3}$.
For the choice of $\sin(\beta-\alpha_1)=0$, $s_3=1$, and $c_2\to 0$, $S$ couples to fermions and gauge bosons as given in Eq.~\eqref{effcoup} with $\theta=\frac{\pi}{2}-\alpha_2$. 
The cubic scalar interactions involving $h$ and $S$ contain
\beq
\mathcal{L}_{\text{cubic}} = - \lambda_{hSS} hSS - \lambda_{Shh} Shh,
\eeq
with
\bea
\lambda_{hSS} &=& \frac{-(\lambda_7  c_\beta^2 + \lambda_8  s_\beta^2) s_2^3 v - (2\lambda_7  c_\beta^2 + 2\lambda_8  s_\beta^2 - 3\lambda_6) s_2^2 c_2 v_s}{2}, \nonumber\\
\lambda_{Shh} &=& \frac{(\lambda_7  c_\beta^2 + \lambda_8  s_\beta^2) s_2^3 v_s 
+ (3\lambda_1 c_\beta^4 + 3\lambda_2 s_\beta^4+6\lambda_{345} s_\beta^2 c_\beta^2-2\lambda_7  c_\beta^2 -2 \lambda_8  s_\beta^2  ) s_2^2 c_2 v}{2},\nonumber\\
\eea
where $\lambda_{345}=\lambda_3+\lambda_4+\lambda_5$, $s_{\beta}\equiv\sin\beta$, and $c_{\beta}\equiv\cos\beta$. 
We retain terms only up to $\order(c_2)$.
Tiny values of $v_s$ and $c_2$ can significantly suppress the coupling $\lambda_{Shh}$.

Besides the two model scenarios discussed above, in the $Y$=0 Higgs triplet model \cite{Gunion:1989ci,Wang:2013jba}, the couplings of the additional CP-even scalar state to fermions arise solely from doublet-triplet mixing, as shown in Eq.~\eqref{effcoup}.
In contrast, its couplings to the gauge bosons receive contributions not only from the mixing angle but also from the VEV of the triplet field. 
In the regime where the additional scalar predominantly decays into fermion pairs, the detection prospects derived from Eq.~\eqref{effcoup} are applicable to the $Y=$0 Higgs triplet model.

In this work, instead of being restricted to any specific model, we work with Eq.~\eqref{effcoup} that can represent a class of models.
Thus, our analysis depends on only three parameters: $\sin\theta$, $m_S$, and $\lambda$, where $\lambda$ is the dimensionless coupling defined in Eq.~\eqref{effcoup}.
Throughout this work, the physical trilinear $hSS$ coupling is normalized as $\lambda v$.
The parameters $\sin\theta$ and $m_S$ determine the decay patterns and lifetime of the scalar $S$, while $m_S$ and $\lambda$ controls the scattering cross section of $gg\to h^* \to SS$ at the LHC.
Owing to the mixing between $S$ and $h$, the $hSS$ coupling is inherently associated with a corresponding 
$Shh$ interaction, as illustrated in the two UV-complete models described above; this coupling can lead to the $S\to h h$ decay if kinematically allowed.
As we focus on a long-lived scalar $S$, once the decay channel $S\to hh$ becomes kinematically accessible, it can have a noticeable impact on the lifetime of $S$, even though the $Shh$ coupling is suppressed.
For simplicity, we choose not to discuss this possibility and restrict our study to the mass range 65 GeV~$<m_S<$~250 GeV, where the lower limit 65~GeV is just above $m_h/2$ and the upper limit 250~GeV corresponds to twice the SM-like Higgs-boson mass.

\section{Experimental analysis}\label{sec:experiment}
The scalar $S$ can decay into $f\bar{f}$, $WW$, $ZZ$, $\gamma\gamma$, and $gg$, and the corresponding partial decay widths are 
\beq
\Gamma_{S\to XX}= \sin^2\theta \times \Gamma_{S\to XX}^{\text{SM}},
\label{eqdecay}
\eeq
where $\Gamma_{S\to XX}^{\text{SM}}$ denotes the partial decay width evaluated with the SM couplings. 
If $\sin\theta$ is sufficient small, $S$ becomes an LLP.
It follows from Eq.~\eqref{eqdecay} that Br$(S\to XX)$ is independent of $\sin\theta$.

\begin{figure}[t]
\begin{center}
 \includegraphics[width=0.4\textwidth]{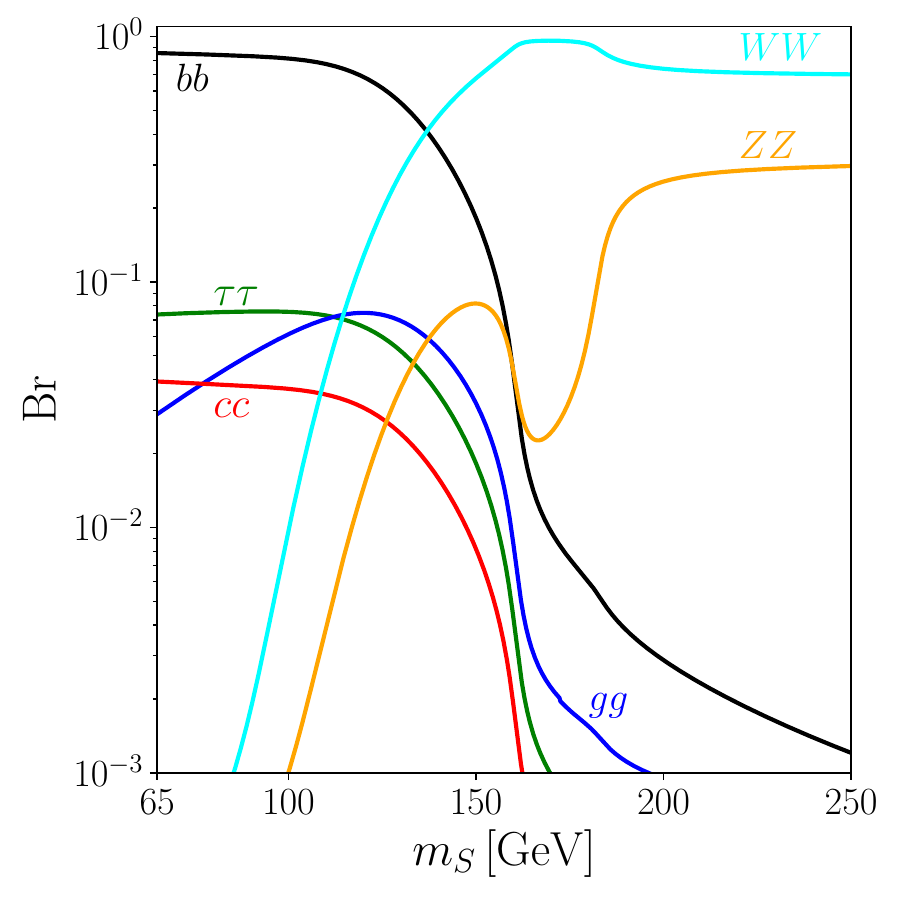}
 \includegraphics[width=0.45\textwidth,height=0.46\textwidth]{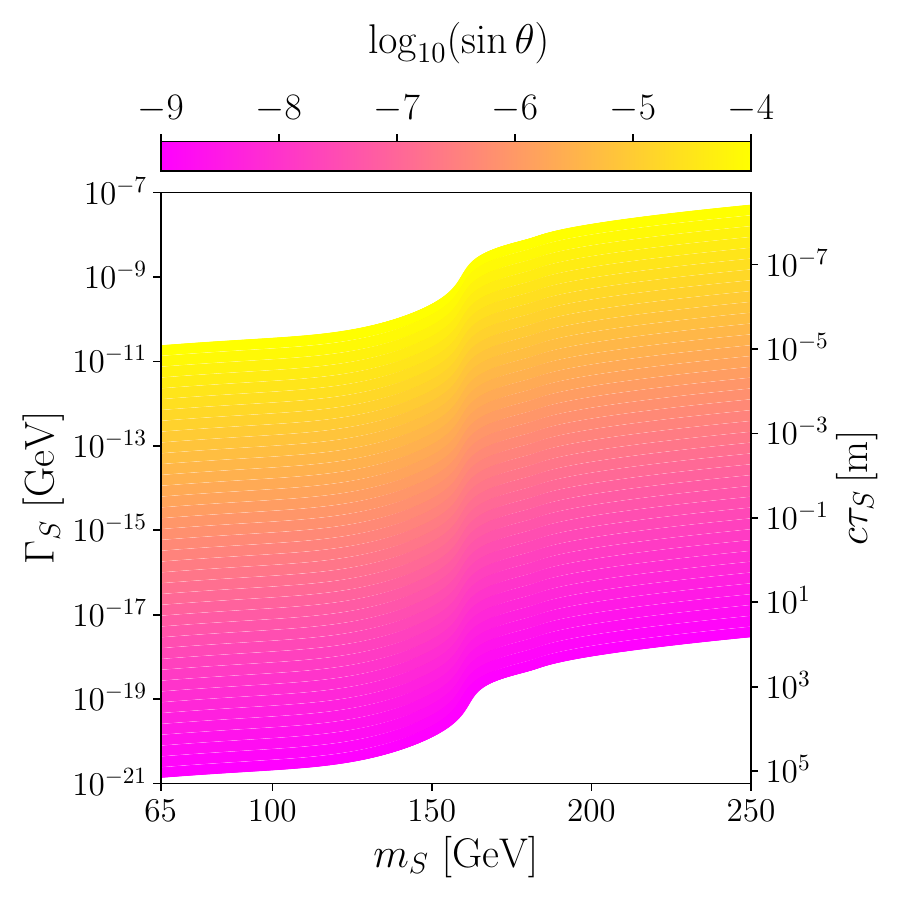}
\end{center}
\vspace{-1.0cm}
\caption{Left panel: the branching ratios of various decay channels of $S$ as functions of its mass. Right panel: the total decay width of $S$ as a function of $m_S$ and $\sin{\theta}$. The corresponding proper decay length $c\tau_S$ is indicated on the right $y$-axis.}
\label{fig:wid}
\end{figure}

We modify the \texttt{2HDMC}~\cite{Eriksson:2009ws} package to compute the decay widths of $S$ with masses between 65~GeV and 250~GeV.
The corresponding branching ratios for various channels are presented in the left panel of figure~\ref{fig:wid}.
For $m_S<$ 140 GeV, the dominant decay mode is $S\to b\bar{b}$.
For a heavier $S$, the $WW$ channel becomes dominant, and the decay $ZZ$ channel can surpass the $b\bar{b}$ one, mainly owing to the opening of these gauge-boson channels and the enhanced couplings to these gauge bosons.

The right panel of figure~\ref{fig:wid} displays the total decay width of $S$, $\Gamma_S$, as a function of $m_S$, for various levels of $\sin\theta$, along with the corresponding proper decay length $c\tau_S$.
$\Gamma_S$ grows with increasing $m_S$ and scales with $\sin^2\theta$.
If $\Gamma_S$ is sufficiently small, the associated decay length $c\tau_S$ becomes macroscopic.

\begin{figure}[t]
\begin{center}
 \includegraphics[width=0.5\textwidth]{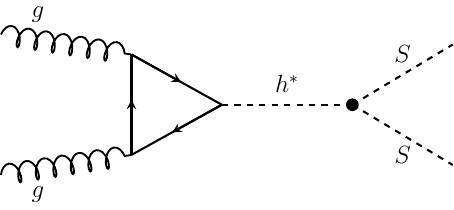}
 \end{center}
\vspace{-1.0cm} \caption{The Feynman diagram for $gg\to h^* \to S S$ at the LHC. The SM-like Higgs boson $h$ is off shell for the $S$-boson masses of our interest and the black blob on the right denotes the vertex associated with the $hSS$ coupling.}
\label{fig:ggh}
\end{figure}

Although the decay widths of the long-lived scalar $S$ is suppressed by the mixing angle $\theta$, the $hSS$ coupling mediating the production of $S$ at the LHC remains unaffected.
Thus, for $m_S>$ 65 GeV, the production of $S$ is dominated by the process $gg\to h^* \to SS$, where the SM-like Higgs boson $h$ is off shell.
The corresponding Feynman diagram is shown in figure~\ref{fig:ggh}.

The long-lived scalar $S$ can travel a macroscopic distance from its production point before decaying inside the inner detector of, for example, ATLAS and CMS.
Depending on its mass, $S$ predominantly decays into a pair of $b$-quarks or a pair of weak gauge bosons ($W^+W^-$ and $ZZ$); the latter case is further dominated by the hadronic decay modes of the $W^{\pm}$- and $Z$-bosons.
As a result, the primary collider signature is characterized with DVs accompanied by several jets at the ATLAS and CMS detectors.

This signature has been searched for at the ATLAS experiment with the full LHC Run-2 dataset of 139~fb$^{-1}$~\cite{ATLAS:2023oti}, and the search targets long-lived electroweakinos in the R-parity-violating supersymmetry decaying into multiple jets with baryon-number-violating operators.
The ATLAS analysis made no discoveries of such electroweakinos and obtained stringent constraints on the masses and lifetimes of these electroweakinos.
Following the ATLAS instructions~\cite{atlas_recast_instruction}, the authors of Ref.~\cite{Cheung:2024qve} have recast the search and reinterpreted the exclusion bounds in a scenario of an axion-like particle with quark-flavor-violating couplings (see also a similar attempt of the recast of the same search in Ref.~\cite{Heisig:2024xbh}).
It is worth noting that the recast has been validated in Ref.~\cite{Cheung:2024qve} for both final states of light-flavor and heavy-flavor jets.
The recast has been applied for further reinterpretation in Refs.~\cite{Wang:2024ieo,Beltran:2025ilg,Wang:2025zhd}, where, in particular, Ref.~\cite{Wang:2024ieo} studied long-lived light scalars produced in \textit{on-shell} SM-like Higgs-boson decays.

Moreover, in Ref.~\cite{Beltran:2025ilg}, the authors proposed a modified analysis based on the same search.
The modified analysis employs lowered jet-$p_T$ thresholds inspired by an 8-TeV ATLAS search for DVs~\cite{ATLAS:2015oan}, while keeping the DV-selection criteria and DV-reconstruction efficiencies from the 13-TeV ATLAS search~\cite{ATLAS:2023oti}.
It is argued that the background levels should remain low or negligible despite the lowered jet-$p_T$ thresholds, mainly because the background suppression is dominantly realized by the vertex-level selections and reconstruction and that the original, relatively high jet-$p_T$ thresholds are imposed in order to ensure sufficient computing resource given the large data-incoming rates at the 13-TeV run.
The modified analysis has been used in Ref.~\cite{Wang:2025zhd}, too, where a long-lived pseudoscalar particle in the type-I 2HDM is studied.
We will apply both analyses to constrain our theoretical scenario of a long-lived scalar produced in pair from off-shell Higgs decays at the LHC, referring to the two analyses as ``original'' and ``modified'' analyses, respectively.

We note that in principle background levels from fake vertices, material interactions, merged low-mass vertices, and QCD multijet events could be raised by the lowered jet-$p_T$ thresholds in the modified analysis; therefore, the sensitivity results of this proposed search should rather be interpreted as projections under optimistic assumptions.

Both the original and modified analyses are executed in two steps: acceptance requirements and parameterized efficiencies, and both steps proceed at both event- and vertex-levels.
The acceptance requirements consist of thresholds on the truth-jets' $p_T$ at the event level and high-quality-DV selection criteria at the vertex level.
The parameterized efficiency functions are provided by the ATLAS collaboration~\cite{atlas_recast_instruction} in order to account for delicate requirements that are difficult to simulate with Monte Carlo (MC) event-generation tools, such as multi-jet triggers and detector-material effects.

In the recast, in order to reconstruct the truth-jets, a toy-detector module has been implemented in \texttt{Pythia8}~\cite{Sjostrand:2014zea,Bierlich:2022pfr}, inspired by the ATLAS recast instructions~\cite{atlas_recast_instruction}.
In detail, we reconstruct the truth-jets with FastJet~\cite{Cacciari:2011ma}, applying the anti-$k_t$ algorithm with $R = 0.4$ where neutrinos and muons are excluded.
Here, the jet definition includes both prompt and displaced jets.
We have, in addition, followed Ref.~\cite{Allanach:2016pam} to model the detector responses to the jet-$p_T$ measurements, taking into account detector acceptance, resolution, and smearing on the truth jets' $p_T$.

The original analysis consists of both a ``high-$p_T$-jet' signal region (SR) and a ``trackless-jet'' one.
In our theoretical scenario the trackless-jet SR shows stronger performance and we will therefore confine ourselves to it.

\begin{table}[t]
\begin{center}
\begin{tabular}{c|c|c}
Analysis            & Original            & Modified                                                              \\ \hline
              & $n^{137}_{\text{jet}}\geq 4$ or $n^{101}_{\text{jet}}\geq 5$    & $n^{90}_{\text{jet}}\geq 4$ or $n^{65}_{\text{jet}}\geq 5$                   \\
jet-$p_T$ thresholds  & or $n^{83}_{\text{jet}}\geq 6$ or $n^{55}_{\text{jet}}\geq 7$, & or $n^{55}_{\text{jet}}\geq 6$                 \\
       &      $n^{70}_{\text{displaced jet}}\geq 1$ or $n^{50}_{\text{displaced jet}}\geq 2$                                                           &  \\ \hline
\end{tabular}
\caption{Event-level acceptance in terms of jet-$p_T$ thresholds. To explain our notation, for example, $n^{137}_{\text{jet}}$ denotes the number of truth jets with a $p_T$ of at least 137 GeV.}
\label{tab:jet_pt_requirement}
\end{center}
\end{table}

Both two analyses start off with jet-$p_T$-threshold selections, listed in detail in Table~\ref{tab:jet_pt_requirement}.
Additionally, in the original analysis, certain numbers of sufficiently hard ``displaced jets'' are required, where a “displaced jet” is defined as a jet determined to have stemmed from the decay of an LLP by requiring a small $\Delta R$ between the LLP’s decay products and the truth jet.

On the events that have passed the jet-$p_T$-threshold requirements, we impose a further set of requirements to ensure the existence of at least one high-quality DV.
In detail, there should be at least one DV from an LLP decay in the event that should fulfill the following criteria:
\begin{enumerate}
    \item Fiducial volume: $4\text{~mm}<R_{xy}<300\text{~mm}$ and $|z|<300\text{~mm}$, where $R_{xy}$ and $|z|$ are the absolute distance of the DV to the IP in the transverse and longitudinal directions, respectively.
    \item Transverse impact parameter: the DV should have at least one associated track with its absolute transverse impact parameter larger than $2$~mm: $|d_0|>2\text{~mm}$.
    \item Selected decay products: the DV should have at least 5 massive decay products passing two conditions:
    \begin{enumerate}
        \item The decay product should be a track with a transverse decay length in the laboratory frame ($\beta_T \gamma c\tau$) longer than 520~mm, where $\beta_T$ is its absolute speed in the transverse direction, $\gamma$ is its Lorentz boost factor, and $c\tau$ is its proper decay length.
        \item The transverse momentum $p_T$ and the electric charge $q$ of the decay product should fulfill $p_T/|q|> 1\text{~GeV}$.
    \end{enumerate}
    \item DV invariant mass:  $m_{\text{DV}}> 10$ GeV, where $m_{\text{DV}}$ is reconstructed with the decay products passing the conditions described above where such decay products are assumed to be a charged pion.
\end{enumerate}

Afterwards, we apply the parameterized efficiency functions mentioned above onto the events that have passed both the event- and vertex-level acceptances described above.
In the original (modified) analysis, we include in the computation both the event- and vertex-level efficiency functions (only the vertex-level efficiency functions).
The event-level parameterized efficiencies are determined by the truth-jet scalar $p_T$ sum and $R_{xy}$ of the furthest LLP decay, and the vertex-level parameterized efficiency functions employ input parameters of $R_{xy}, m_{\text{DV}}$, as well as LLP decay-product multiplicity.

Potential backgrounds dominantly originate from erroneous merge of nearby DVs of small
invariant masses by vertexing algorithms leading to a large $m_{\text{DV}}$, particle hadronic interactions with detector materials, and accidental crossings between a track and unrelated low-mass DVs.
After the event-selection criteria given above are applied, almost vanishing backgrounds are expected.
Concretely, the ATLAS analysis estimated the background-event yields to be $0.83^{+0.51}_{-0.53}$ in this SR and observed 0 events in the data.

Furthermore, we make projections for the HL-LHC with an integrated luminosity of $3$~ab$^{-1}$, where we assume the backgrounds can be contained at the similarly vanishingly low levels, in expectation of developments in accelerator and detector technologies as well as analysis algorithms.

\section{Numerical results}\label{sec:results}
We briefly outline our simulation setup.
We implement the phenomenological model (Eq.~\eqref{effcoup}) in \texttt{FeynRules}~\cite{Christensen:2008py,Alloul:2013bka} and \texttt{NLOCT} packages \cite{Degrande:2014vpa} and obtain the corresponding UFO model files, which are then interfaced with \texttt{MadGraph5aMC$@$NLO}~\cite{Alwall:2014hca,Frederix:2018nkq} to generate leading-order (LO) parton-level signal events.
The cross section of the process $gg \to h^* \to SS$ is calculated using the exact one-loop amplitude, retaining the complete dependence on $m_t$.
No heavy-top effective field theory approximation is employed in the generation of the LO matrix elements. 
The renormalization and factorization scales are chosen as $\mu_R=\mu_F=m_{SS}$, where $m_{SS}$ denotes the invariant mass of the $SS$ system.
Regarding higher-order QCD effects, the next-to-leading-order (NLO) corrections to gluon-fusion processes are generally not uniform over the entire $m_{SS}$ range.
However, the primary goal of the present work is not to provide a precision prediction of the inclusive production rate, but rather to investigate the collider sensitivity to the long-lived scalar particles.
Based on the results of Ref.~\cite{Borowka:2016ypz,Baglio:2018lrj,Hespel:2014sla}, we adopt a constant $K$-factor of $1.6$ as an effective approximation to account for the dominant NLO QCD enhancement.
Compared with the exact $m_{SS}$-dependent $K$-factor, we estimate that the adopted $K$-factor may introduce deviations of up to approximately $20\%$ in certain regions of $m_{SS}$.
Nevertheless, this treatment is expected to provide a more realistic estimate of the signal rates than a purely LO prediction and is sufficient for the phenomenological sensitivity study presented in this work.

The resulting LHE event samples~\cite{Alwall:2006yp} are subsequently passed to \texttt{Pythia8} for parton showering and hadronization.
In the simulation, \texttt{Pythia8} handles all the decay channels of $S$ according to their respective branching ratios.
We then apply our recast framework to evaluate the cutflow efficiencies over the scanned parameter space.

The number of signal events is computed with
\bea
N_S=\mathcal{L}_\text{int.}\cdot \sigma \cdot \epsilon.
\eea
Here, $\mathcal{L}_\text{int.}=139~\text{fb}^{-1}$ or $3000~\text{fb}^{-1}$ denotes the integrated luminosity, $\sigma$ is the total signal cross section, and $\epsilon$ represents the overall selection efficiency obtained from our recast analysis.
For the original search analysis at the LHC with $\mathcal{L}_\text{int.}=139\text{~fb}^{-1}$, signal events are generated at the center-of-mass energy $\sqrt{s}=$ 13~TeV, while for all other analyses, signal samples are generated at $\sqrt{s}=$ 14~TeV.
At each parameter point, at least one million events are simulated so that it is ensured that the cutflow efficiencies should have been stabilized.

\begin{figure}[t]
\begin{center}
\includegraphics[width=0.5\textwidth]{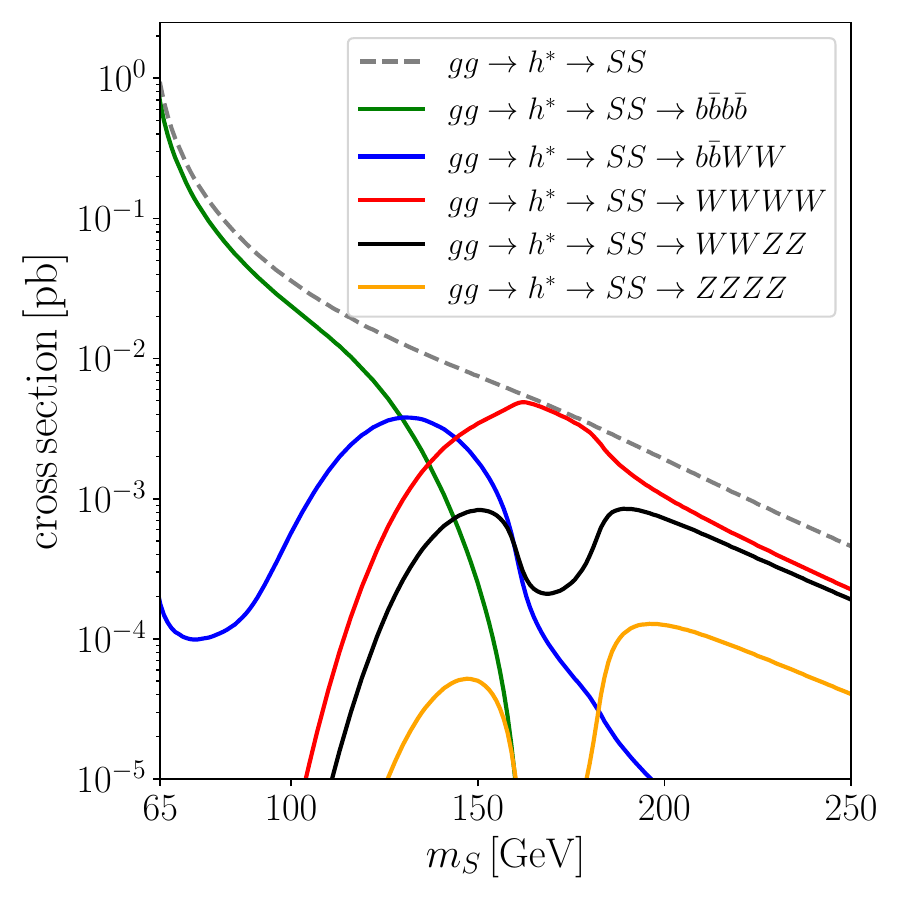}
\end{center}
\vspace{-1.0cm} \caption{Cross sections of $gg\to h^*\to SS$ and $gg\to h^*\to SS\to XXYY$ where $XXYY$ includes several main decay channels of $S$ in the mass range of interest, as functions of $m_S$, at the 14~TeV LHC, assuming $\lambda=1.0$.}
\label{fig:cs}
\end{figure}

In Figure~\ref{fig:cs}, we fix the coupling parameter $\lambda=1.0$, and displays the \texttt{MadGraph5}-computed cross sections of the signal processes as functions of $m_S$ at $\sqrt{s}=14$~TeV, for $gg\to h^*\to SS$ and $gg\to h^*\to SS\to XXYY$ where $XXYY$ denotes several main decay channels of the two $S$'s particles in the mass range of interest.
In our simulation, no specific decay mode of $S$ is imposed.
The individual final states shown here are for illustration only and serve to indicate the relative importance of different decay channels.
The total cross section of $gg\to h^*\to SS$ decreases with $m_S$, primarily owing to phase-space suppression.
As $m_S$ increases, the dominant final state changes from $b\bar{b}b\bar{b}$ to $b\bar{b}WW$, and eventually to $WWWW$.
This behavior is in good agreement with the decay branching ratios of $S$ shown in the left panel of figure~\ref{fig:wid}.

It is worth noting that the fixed value $\lambda=1.0$ lies well within the perturbative regime. For example, in the real-singlet scalar extension of the SM, $\lambda$ may be interpreted as the quartic coupling $\lambda_5$. Perturbative unitarity typically requires \cite{Kanemura:2015fra,He:2016sqr}
\begin{eqnarray}  
&&| \lambda_1| < 8\pi,~ | \lambda_5 | < 8\pi,\nonumber\\
&&\frac{1}{2}\left| \frac{3}{2}\lambda_2+3\lambda_1\pm \sqrt{4\lambda_5^2+(\frac{3}{2}\lambda_2-3\lambda_1)^2}\right| < 8\pi.
\end{eqnarray} 
Therefore, $\lambda_5=1$ is far from the corresponding perturbative limit. 
In addition, the scalar potential in Eq.~\eqref{VSMS} contains a quartic interaction $hhSS$ with the coupling strength $g_{hhSS}=\mathcal{O}(\lambda_5$).
For $\lambda_5=1$, this coupling is also substantially smaller than the characteristic strong-coupling scale $\mathcal{O}(4\pi)$.
We thus conclude that the benchmark value $\lambda=1$ is not expected to indicate a breakdown of perturbation theory.

\begin{table}[t]
\begin{center}
\begin{tabular}{|c|c|c|c|}

\hline
$m_S$~[GeV]        &cross section [pb]  & Scale uncertainty (\%) &  PDF uncertainty (\%)\\
\hline
100                &    $3.6\times 10^{-2}$           & +23/-18                &  $\pm$ 1.64 \\
\hline
170                &     $4.5\times 10^{-3}$         & +27/-20                &  $\pm$ 1.68 \\
\hline
250               &    $4.6\times 10^{-4}$         & +29/-21                &  $\pm$ 1.88 \\
\hline

\end{tabular}

\caption{Scale and PDF uncertainties of the signal production cross section for three representative benchmark points with the coupling parameter fixed at $\lambda=1.0$.}
\label{tab:uncertain}
\end{center}
\end{table}
The scale uncertainty on the signal cross section is evaluated by varying the renormalization and factorization scales by a factor of two around the central value.
The PDF uncertainty is evaluated using the \texttt{NNPDF30} error set through the LHAPDF framework~\cite{NNPDF:2014otw,Buckley:2014ana}.
To explicitly demonstrate the impact of these theoretical uncertainties, we display the scale and PDF uncertainties for three representative benchmark points, $m_S=100$~GeV, 170~GeV, and 250~GeV, in Table~\ref{tab:uncertain}.
We find that the signal cross sections exhibit scale uncertainties at the level of approximately $20\%-30\%$, while the PDF uncertainties remain at the percent level.
Therefore, the theoretical uncertainty is dominated by the scale variation, whereas the PDF uncertainty has only a minor impact on the signal prediction.

\begin{table}[htbp!]
\begin{center}
\resizebox{\textwidth}{!}{
\begin{tabular}{l|cc|cc|cc}

\hline
$m_S$~[GeV], $\sin{\theta}$, $c\tau_S$~[m]        & \multicolumn{2}{c|}{$100, 10^{-8}, 7.2 \times 10^{2}$} & \multicolumn{2}{c|}{$100, 10^{-7}, ~7.2$ } & \multicolumn{2}{c}{$100, 5\times 10^{-7}, 2.9 \times 10^{-1}$} \\
\hline
Jet-$p_T$ threshold                            &    $4.7\times 10^{-5}$ & $4.1\times 10^{-2}$ &     $3.5\times 10^{-3}$  & $4.0\times 10^{-2}$ & $9.0\times 10^{-3}$   & $4.0\times 10^{-2}$ \\
Event has $\geq 1$ DV passing:                 &                        &                     &                          &                     &                       &                            \\
~~$R_{xy}, |z|<300$ mm               &          $4.4\times 10^{-6}$     & $2.3\times 10^{-5}$ &     $4.9\times 10^{-4}$  & $2.5\times 10^{-3}$ & $6.2\times 10^{-3}$   & $2.8\times 10^{-2}$    \\
~~$R_{xy}>4$ mm                      &          $4.4\times 10^{-6}$     & $2.3\times 10^{-5}$ &     $4.8\times 10^{-4}$  & $2.4\times 10^{-3}$ & $6.2\times 10^{-3}$   & $2.8\times 10^{-2}$   \\
~~$\geq 1$ trk.~with $|d_0| > 2$ mm &           $4.4\times 10^{-6}$     & $2.3\times 10^{-5}$ &     $4.8\times 10^{-4}$  & $2.4\times 10^{-3}$ & $6.2\times 10^{-3}$   & $2.8\times 10^{-2}$  \\
~~$n_{\text{sel.~dec.~prod.}}>5$     &          $4.4\times 10^{-6}$     & $2.1\times 10^{-5}$ &     $4.6\times 10^{-4}$  & $2.3\times 10^{-3}$ & $6.0\times 10^{-3}$   & $2.7\times 10^{-2}$  \\
~~$m_{\text{DV}}>10$ GeV             &          $4.4\times 10^{-6}$     & $2.1\times 10^{-5}$ &     $4.6\times 10^{-4}$  & $2.3\times 10^{-3}$ & $6.0\times 10^{-3}$   & $2.7\times 10^{-2}$    \\
\hline
Param.~Effi.                              &      $2.0\times 10^{-7}$    & $5.0\times 10^{-6}$ &     $3.6\times 10^{-5}$  & $8.4\times 10^{-4}$ & $2.1\times 10^{-3}$   & $1.5\times 10^{-2}$   \\
\hline

\hline
$m_S$~[GeV], $\sin{\theta}$, $c\tau_S$~[m]        & \multicolumn{2}{c|}{$170, 10^{-8}, ~5.5$} & \multicolumn{2}{c|}{$170, 10^{-7}, 5.5 \times 10^{-2}$} & \multicolumn{2}{c}{$170, 5\times 10^{-7}, 2.2 \times 10^{-3}$} \\
\hline
Jet-$p_T$ threshold                            &    $1.1\times 10^{-2}$ & $6.3\times 10^{-2}$ &     $1.8\times 10^{-2}$  & $6.3\times 10^{-2}$ & $1.8\times 10^{-2}$   & $6.3\times 10^{-2}$  \\
Event has $\geq 1$ DV passing:                 &                        &                     &                          &                     &                       &                       \\
~~$R_{xy}, |z|<300$ mm               &          $1.7\times 10^{-3}$     & $6.7\times 10^{-3}$ &     $1.8\times 10^{-2}$  & $6.3\times 10^{-2}$ & $1.8\times 10^{-2}$   & $6.3\times 10^{-2}$   \\
~~$R_{xy}>4$ mm                      &          $1.7\times 10^{-3}$     & $6.6\times 10^{-3}$ &     $1.8\times 10^{-2}$  & $6.2\times 10^{-2}$ & $1.0\times 10^{-2}$   & $3.1\times 10^{-2}$    \\
~~$\geq 1$ trk.~with $|d_0| > 2$ mm &           $1.7\times 10^{-3}$     & $6.6\times 10^{-3}$ &     $1.8\times 10^{-2}$  & $6.2\times 10^{-2}$ & $9.9\times 10^{-3}$   & $3.0\times 10^{-2}$   \\
~~$n_{\text{sel.~dec.~prod.}}>5$     &          $1.7\times 10^{-3}$     & $6.2\times 10^{-3}$ &     $1.8\times 10^{-2}$  & $6.1\times 10^{-2}$ & $9.8\times 10^{-3}$   & $3.0\times 10^{-2}$   \\
~~$m_{\text{DV}}>10$ GeV             &          $1.6\times 10^{-3}$     & $6.2\times 10^{-3}$ &     $1.8\times 10^{-2}$  & $6.1\times 10^{-2}$ & $9.8\times 10^{-3}$   & $3.0\times 10^{-2}$   \\
\hline
Param.~Effi.                              &      $1.9\times 10^{-4}$    & $2.6\times 10^{-3}$ &     $1.2\times 10^{-2}$  & $5.5\times 10^{-2}$ & $6.7\times 10^{-3}$   & $2.8\times 10^{-2}$ \\
\hline

\hline
$m_S$~[GeV], $\sin{\theta}$, $c\tau_S$~[m]        & \multicolumn{2}{c|}{$250, 10^{-8}, 5.1 \times 10^{-1}$} & \multicolumn{2}{c|}{$250, 10^{-7}, 5.1 \times 10^{-3}$} & \multicolumn{2}{c}{$250, 5\times 10^{-7}, 2.0 \times 10^{-4}$} \\
\hline
Jet-$p_T$ threshold                            &    $5.5\times 10^{-2}$ & $1.7\times 10^{-1}$ &     $5.5\times 10^{-2}$  & $1.7\times 10^{-1}$ & $5.5\times 10^{-2}$   & $1.7\times 10^{-1}$    \\
Event has $\geq 1$ DV passing:                 &                        &                     &                          &                     &                       &                                \\
~~$R_{xy}, |z|<300$ mm               &          $3.8\times 10^{-2}$     & $1.2\times 10^{-1}$ &     $5.5\times 10^{-2}$  & $1.7\times 10^{-1}$ & $5.5\times 10^{-2}$   & $1.7\times 10^{-1}$   \\
~~$R_{xy}>4$ mm                      &          $3.7\times 10^{-2}$     & $1.2\times 10^{-1}$ &     $4.1\times 10^{-2}$  & $1.1\times 10^{-1}$ & $1.6\times 10^{-5}$   & $2.0\times 10^{-5}$    \\
~~$\geq 1$ trk.~with $|d_0| > 2$ mm &           $3.7\times 10^{-2}$     & $1.2\times 10^{-1}$ &     $4.1\times 10^{-2}$  & $1.1\times 10^{-1}$ & $1.4\times 10^{-5}$   & $1.7\times 10^{-5}$    \\
~~$n_{\text{sel.~dec.~prod.}}>5$     &          $3.7\times 10^{-2}$     & $1.2\times 10^{-1}$ &     $4.0\times 10^{-2}$  & $1.1\times 10^{-1}$ & $1.4\times 10^{-5}$   & $1.7\times 10^{-5}$  \\
~~$m_{\text{DV}}>10$ GeV             &          $3.7\times 10^{-2}$     & $1.2\times 10^{-1}$ &     $4.0\times 10^{-2}$  & $1.1\times 10^{-1}$ & $1.4\times 10^{-5}$   & $1.7\times 10^{-5}$   \\
\hline

Param.~Effi.                              &      $1.7\times 10^{-2}$    & $7.7\times 10^{-2}$ &     $2.9\times 10^{-2}$  & $1.1\times 10^{-1}$ & $2.4\times 10^{-6}$   & $1.7\times 10^{-5}$ \\
\hline

\end{tabular}
}
\caption{Cutflow efficiencies obtained from $10^6$ MC events for benchmark values of $m_S$ and $\sin\theta$.
For the benchmark points where acceptances are below $\mathcal{O}(10^{-5})$ more signal events are generated to ensure stabilized efficiencies.
The corresponding $c\tau_S$ values are computed with the modified \texttt{2HDMC} package. For each benchmark point, two sets of efficiencies are listed, for the original and modified analyses, respectively.
Compared to the original analysis, the modified one adopts looser jet-$p_T$ cuts and does not employ the event-level parameterized efficiency functions.}
\label{tab:cutflow}
\end{center}
\end{table}

Table~\ref{tab:cutflow} summarizes the cutflow efficiencies obtained from $10^6$ simulated signal events at multiple selected benchmark points.
Each benchmark point is characterized by a specific combination of $m_S$ and $\sin\theta$, together with the corresponding value of $c\tau_S$ computed with the \texttt{2HDMC} package.
We have chosen representative values of $m_S=100$~GeV, 170~GeV, 250~GeV, and $\sin\theta=10^{-8},10^{-7}, 5\times 10^{-7}$.
We emphasize that the parameter $\lambda$ affects only the cross section and has no impact on the cutflow efficiencies.
The cutflow efficiencies obtained from the original and modified analyses are listed side by side for comparison, showing consistently improved final efficiencies in the modified analysis across all benchmark points.

For any fixed value of $m_S$, increasing $\sin\theta$ and thus decreasing $c\tau_S$ can reflect a transition behavior of the scalar $S$ from tending to decay outside the fiducial volume, to decay inside the fiducial volume, and finally to decay promptly.
Also, for similar values of $c\tau_S$, a benchmark point of a heavier $S$ leads to larger cutflow efficiencies; this is mainly due to the larger probabilities to pass the jet-$p_T$ thresholds and the high-quality DV selections.
The final cutflow efficiencies listed in the table~\ref{tab:cutflow} range from $\mathcal{O}(10^{-7})$ to $\mathcal{O}(10^{-1})$.

\begin{figure}[t]
\begin{center}
 \includegraphics[width=0.45\textwidth]{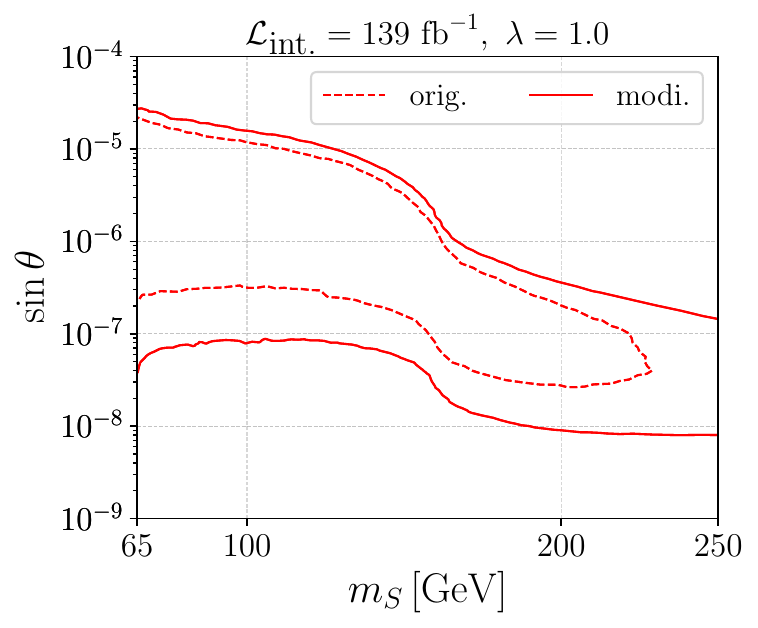}
 \includegraphics[width=0.45\textwidth]{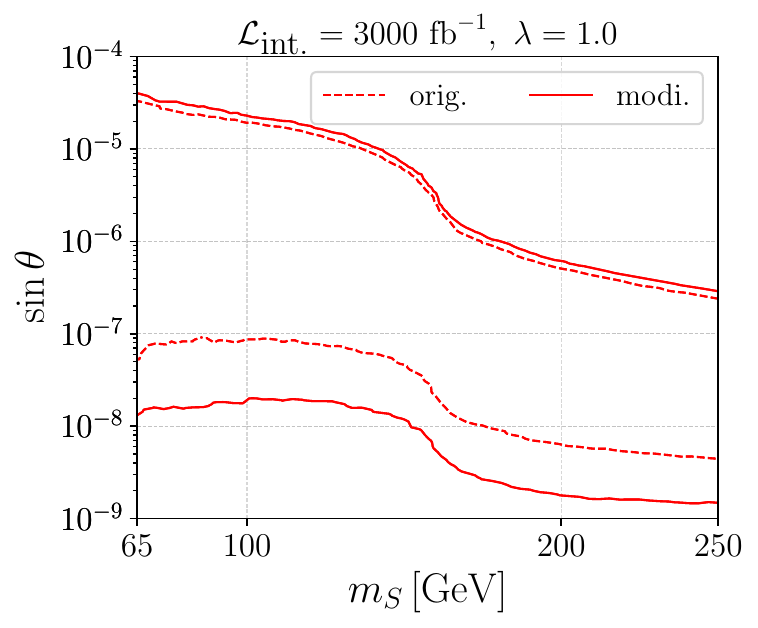}
 \end{center}
\vspace{-1.0cm} \caption{Sensitivity reach to $\sin\theta$ at $95\%$ C.L.~as a function of $m_S$, for $\lambda=1.0$. The left (right) panel shows the results for an integrated luminosity of $\mathcal{L}_{\text{int.}}=139$~fb$^{-1}$ ($\mathcal{L}_{\text{int.}}=3000$~fb$^{-1}$). The dashed and solid lines correspond to the original and modified analyses, respectively.}
\label{fig:mssin}
\end{figure}

For both analyses, the expected background is negligible.
We therefore adopt 3 signal events as the exclusion criterion at 95\% confidence level (C.L.).
In figure~\ref{fig:mssin}, we fix $\lambda=1.0$, and show the exclusion limits in the $\sin\theta$~vs.~$m_S$ plane for integrated luminosities of 139 fb$^{-1}$ (the left panel) and 3000 fb$^{-1}$ (the right panel).

The dependence of the sensitivities on $m_S$ exhibits two distinct features.
As $m_S$ increases, especially above $\sim 165$~GeV, the decay width of $S$ is enhanced owing to the opening of gauge-boson decay modes, thus improving the lower limits to $\sin\theta$ where the scalar particle $S$ is very long-lived and the signal-event rates are approximately proportional to $\Gamma_S$.
At the same time, the production cross sections decrease with $m_S$, thus reducing
the overall sensitivity and narrowing the excluded region.
In particular, for the original analysis with an integrated luminosity of 139~fb$^{-1}$ (the dashed line in the left panel), the upper reach to $m_S$ is about $230$~GeV.

\begin{figure}[t]
\begin{center}
 \includegraphics[width=0.9\textwidth]{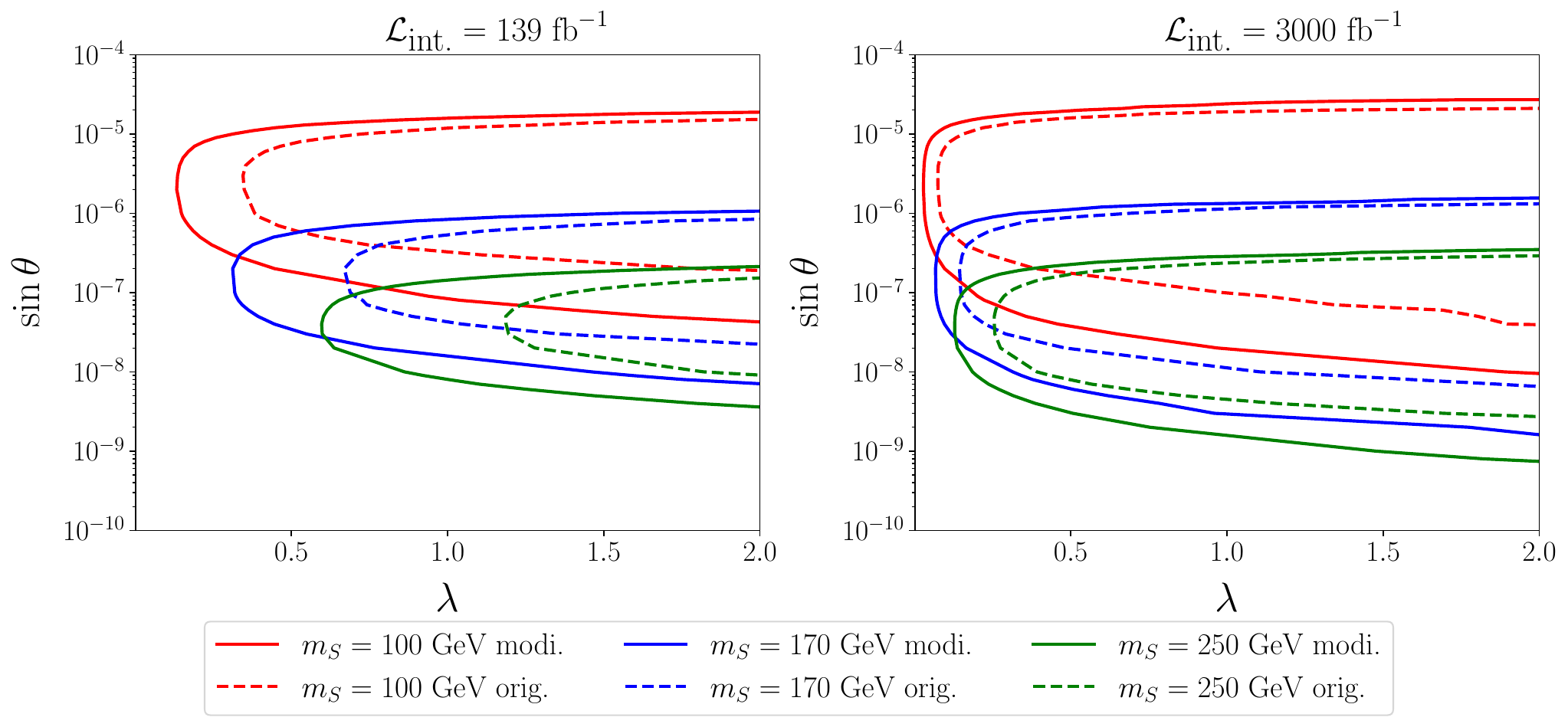}
 \end{center}
\vspace{-1.0cm}  \caption{Sensitivity reach to $\sin\theta$ at $95\%$ C.L., with respect to $\lambda$, for representative values of $m_S$: 100~GeV (red), 170~GeV (blue), and 250~GeV (green). The dashed and solid lines correspond to the original and modified analyses, respectively, and the two plots are for integrated luminosities of 139~fb$^{-1}$ and 3000~fb$^{-1}$, respectively.}
\label{fig:lamsin}
\end{figure}

In figure~\ref{fig:lamsin}, we present the 95\% C.L.~exclusion limits in the $\sin\theta$~vs.~$\lambda$ plane for several representative values of $m_S$: 100~GeV, 170~GeV, and 250~GeV.
The exclusion contours exhibit a clear dependence on $m_S$. 
For heavier $S$, the excluded region shifts toward smaller values of $\sin\theta$ and larger values of $\lambda$. 
This behavior can be understood from the fact that a larger $\lambda$ enhances the production cross section and a smaller $\sin\theta$ renders $S$ more long-lived, considering that a heavier $S$ leads to reduced cross sections and enhances decay widths.
Also, as $m_S$ increases, the area of the overall excluded region in both $\sin\theta$ and $\lambda$ shrinks. 
For example, at an integrated luminosity of 139~fb$^{-1}$ with the original analysis, the exclusion approximately corresponds to $\lambda > 0.35$ for $m_S=100$~GeV, $\lambda > 0.67$ for $m_S=170$~GeV, and $\lambda > 1.2$ for $m_S=250$~GeV.

Both figure~\ref{fig:mssin} and figure~\ref{fig:lamsin} show that the modified analysis consistently outperforms the original analysis for all benchmark points and integrated luminosities. 
In addition, increasing the luminosity from 139~fb$^{-1}$ to 3000~fb$^{-1}$ significantly extends the sensitivity reaches.

It should be noted that the present analysis is intended as a simplified-model study of a long-lived scalar characterized by the phenomenological parameters $m_S$, $s_\theta$, and $\lambda$.
The limits obtained in this work are therefore applicable to a broad class of new-physics scenarios.
When these parameters are interpreted within a specific model, such as the real-singlet scalar extension of the SM, additional theoretical constraints, including perturbative unitarity, boundedness from below, and the global-minimum condition, must be taken into account.
Combined with such theoretical constraints, the collider sensitivities derived in this work can further restrict the parameter space of the underlying model.

\section{Summary}\label{sec:summary}
In this work, we have investigated the prospects of probing a long-lived scalar $S$ through the process $gg \to h^* \to SS$ where the intermediate SM-like Higgs boson $h$ is off shell.
Compared with the conventional on-shell Higgs decay $h\to SS$, this mechanism provides a complementary probe of long-lived $S$ in the mass region above $m_h/2$, thereby significantly extending the accessible parameter space.

We have performed MC simulations of signal events featuring DVs accompanied by multiple jets.
Based on a recast of an ATLAS search for DVs plus jets, together with an earlier 8-TeV ATLAS DV analysis, we evaluate the LHC sensitivity to $S$ with two search strategies (dubbed original and modified analyses).
The study is carried out for integrated luminosities of $139~\text{fb}^{-1}$ and $3000~\text{fb}^{-1}$.
Our results indicate that the DV+jets signatures offer a powerful probe of the long-lived scalar $S$ that couples to SM fermions and gauge bosons via mixing with $h$.
In particular, the existing ATLAS Run-2 analysis already excludes a substantial region of the parameter space for $65\text{~GeV}<m_S<230\text{~GeV}$, assuming $\lambda=1.0$.  
The modified analysis and future upgrades in the HL-LHC era can effectively allow to probe broader regions.

\section*{Acknowledgments} 
This work was supported by the National Natural Science Foundation
of China under grants No.~11975013, 11805001, 12475106 and 12505120, and by the Projects No.~ZR2024MA001 and No.~ZR2023MA038 supported by Shandong Provincial Natural Science Foundation, and the Fundamental Research Funds for the Central Universities under Grant No.~JZ2025HGTG0252.

\vspace{1cm}

\bibliography{refs}

\end{document}